\documentclass{article}

\usepackage{arxiv}

\usepackage[utf8]{inputenc} 
\usepackage[T1]{fontenc}    
\usepackage{hyperref}       
\usepackage{url}            
\usepackage{booktabs}       
\usepackage{amsfonts}       
\usepackage{nicefrac}       
\usepackage{microtype}      
\usepackage{lipsum}		
\usepackage{graphicx}
\usepackage{natbib}
\usepackage{doi}
\usepackage{amsmath}
\usepackage{amssymb}

\title{QED Effects at Grazing Incidence on Solid-State-Targets}

\author{ \href{https://orcid.org/0000-0002-5590-0095}{\hspace{1mm}Marko Filipovic} \\
	Institut f\"ur Theoretische Physik I\\
    Heinrich-Heine-Universit\"at D\"usseldorf\\
	D\"usseldorf, 40225 \\
	\texttt{marko.filipovic@hhu.de} \\
	\And
	\href{https://orcid.org/0000-0001-5043-960X}{\hspace{1mm}Alexander Pukhov} \\
	Institut f\"ur Theoretische Physik I\\
    Heinrich-Heine-Universit\"at D\"usseldorf\\
	D\"usseldorf, 40225 \\
	\texttt{pukhov@tp1.uni-duesseldorf.de} \\
}

\renewcommand{\undertitle}{Preprint submitted to EPJ D}

\hypersetup{
pdftitle={QED Effects at Grazing Incidence},
pdfsubject={physics.plasm-ph},
pdfauthor={Marko Filipovic, Alexander Pukhov},
pdfkeywords={particle-in-cell simulation, QED, photon emission, pair production},
}
\graphicspath{ {pictures/} }

\begin{document}
\maketitle

\begin{abstract}
New laser facilities will reach intensities of $10^{23} \textrm{W cm}^{-2}$. This advance enables novel experimental setups in the study of laser-plasma interaction. In these setups with extreme fields quantum electrodynamic (QED) effects like photon emission via non-linear Compton scattering and Breit-Wheeler pair production become important.    

We study high-intensity lasers grazing the surface of a solid-state target by two-dimensional particle-in-cell simulations with QED effects included. The two laser beams collide at the target surface at a grazing angle. Due to the fields near the target surface electrons are extracted and accelerated. Finally, the extracted electrons collide with the counter-propagating laser, which triggers many QED effects and leads to a QED cascade under a sufficient laser intensity. Here, the processes are studied for various laser intensities and angles of incidence and finally compared to a seeded vacuum cascade. Our results show that the proposed target can yield many order of magnitude more secondary particles and develop a QED cascade at lower laser intensities than the seeded vacuum alone.
\end{abstract}


\section{Introduction}
\label{sec:introduction}
With the construction of high-intensity laser facilities like ELI~\citep{whitebook_ELI},XCELS~\citep{whitebook_XCELS} and Apollon \cite{HighPowerLaserSciEng_7_e54} the study of laser-matter interaction in strong electromagnetic fields has been greatly advanced. The development of these lasers is possible due to the chirped pulse amplification \citep{STRICKLAND1985447} which revolutionized the high-intensity laser technology. One of the natural applications for such lasers is the high-energy electron acceleration that can in turn be used to study fundamental quantum physics. 

Two of the QED effects that can take place when relativistic electrons interact with a strong electromagnetic field are the emission of hard photons by non-linear Compton scattering and electron-positron pair production with the multiphoton Breit-Wheeler process \cite{di2012extremely, narozhny2015quantum,https://doi.org/10.48550/arxiv.2203.00019}. In order to quantify whether the mentioned interactions take place the quantum nonlinearity parameter $\chi$ hasa been defined and reads
\begin{equation}
\chi = \sqrt{-\left(F_{\mu \nu} p^{\left(\nu \right) } \right)^2} / \left( m_e c E_{cr} \right)
\end{equation}
where $E_{cr}$ is the critical field of vacuum breakdown also known as the Schwinger limit \citep{Schwinger1951}; $F_{ \mu \nu }$ the electromagnetic field tensor; and $p^{ \left( \nu \right)}$ being the four-dimensional momentum. The quantum parameter $\chi$  defines whether the processes are treated in classical electrodynamics or quantum electrodynamics and is often used in computational tools to calculate probabilities for QED effects \citep{elkina2011qed}. Once $\chi \geq 1$ the radiation processes should be treated in quantum electrodynamics. Photon emission and pair production may continue repetitively in a strong electromagnetic field and lead to QED cascades~\citep{elkina2011qed,PhysRevLett.105.080402,Bell2008,kirk2009pair,PhysRevE.95.023210}. This may result in an electron-positron plasma of high density. Electrons and positrons oscillating in the strong electromagnetic fields emit photons, while the new photons decay again to an electron-positron plasma, which have the possibility again to repeat the cycle.

The $\chi$-parameter also classifies different QED regions besides distinguishing how to treat processes. The supercritical regime corresponds to a limit of $\chi \gg 1$ \cite{PhysRevD.104.L091903} and after reaching $\alpha \chi^{2/3} \geq 1$ it is conjectured that QED theory becomes nonperturbative \citep{PhysRevD_21_1176, AnnPhys_69_555}. Here, $\alpha$ denotes the fine structure constant. This fully nonperturbative QED (FNQED) regime is still not experimentally explored, but various initial analytic studies were conducted \cite{JPhysConfSer_826_012027, mironov2020resummation, ekman2020high}. 

A second important parameter for QED studies which describes whether a field is able to accelerate an electron to relativistic energies is

\begin{equation}
a_0 =\frac{ \lvert e \rvert E_L }{m_e c \omega_0}
\end{equation}
the dimensionless field amplitude \citep{di2012extremely}. Previously mentioned lasers-facilities will be able to provide the necessary parameter of $a_0 \gg 1$ to witness relativistic particles and observe QED effects within promising configurations if the quantum parameter $\chi$ is big enough.
 
The challenge to reach the new QED regimes is to generate strong fields and high-energetic particles at the same time. These allow particles to achieve a high quantum parameter. Possible setups to study QED effects are the collision between an electron beam and a laser pulse with an intensity of $10^{24} W cm^{-2}$ \citep{NewJPhys_21_053040}, collision of high-current $100 \textrm{GeV}$ electron bunches\cite{PhysRevLett_122_190404}, collision of an ultra-relativistic electron beam with a counter-propagating ultraintense electromagnetic pulse \cite{SciRep_9_9407} and other configurations \citep{PlasmaPhysControlFusion_61_074010,filipovic2021effect,di2020testing}.

Numerical particle-in-cell (PIC) simulations which research particle dynamics of laser-plasma interactions are an important tool to study QED effects \cite{pukhov1999three,10.1007/3-540-47789-6_36,DEROUILLAT2018351}. These codes can include QED effects to simulate future experiments. Here, we use the PIC code \texttt{VLPL} \citep{CERNYellowReports_1_181} with the QED module \citep{elkina2011qed,Baumann2016} to study the proposed configuration in regards of non-linear Compton scattering and multi-photon Breit-Wheeler pair production. 

The work in this paper focuses on the effect of high-intensity lasers at grazing incidence to study QED effects and QED cascades on a solid density target. The PIC code simulates not only charged particles (electrons, positrons, ions), but also treats $\gamma$-photons as numerical particles. This allows us to generate $\gamma$-photons by initializing a laser that propagates towards a solid-state target to extract and accelerate electrons and emit photons by the non-linear Compton scattering. Therefore, the proposed configuration uses two high-intensity lasers that collide on the surface of the target. Using a small grazing incidence achieves a higher Lorentz-factor for the extracted particles and a greater current of the extracted electron bunch \citep{doi:10.1063/1.5002671}.

The paper is organized as follows. In section \ref{sec:setup}, we describe the simulation configuration with the parameters of the domain, target and laser beams. A brief summary of the used QED module will be given as well. Section \ref{sec:results} presents the results of particle-in-cell simulations and the generation of electron-positron plasma in this configuration. We compare the electron-positron plasma production near a solid density surface with the vacuum case. Section \ref{sec:conclusion} summarizes the main results of our study and gives a prospect to the future of this subject.
\section{Simulation setup}
\label{sec:setup}
\begin{figure}
\centering
\includegraphics[scale=0.25]{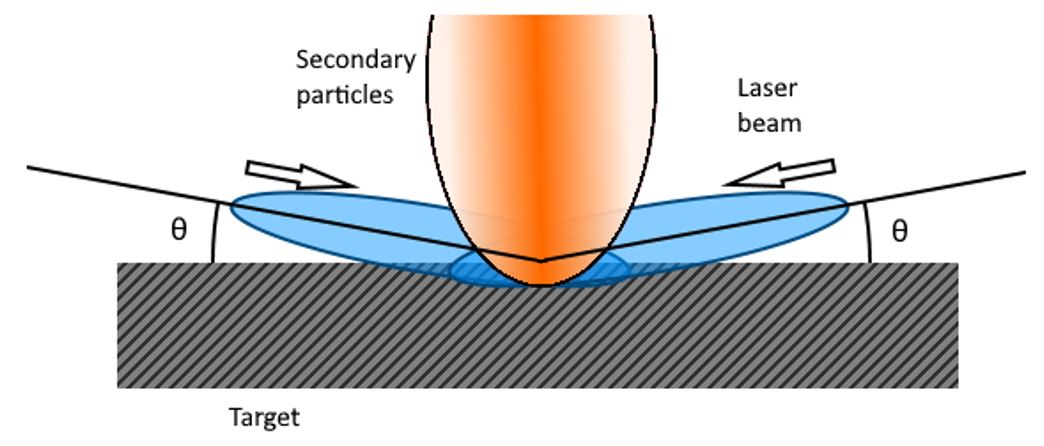}
\caption{ Configuration of two high-intensity lasers grazing a solid-state target. Arrows indicates the trajectory of each laser beam. The lasers are focused on the centre of the upper edge of the target. Both lasers are incident on the same angle $\Theta$. The orange half ellipsis shows the region where secondary particles from QED effects will be located after the interaction.}
\label{fig:1_config}
\end{figure}

The particle-in-cell (PIC) simulations are performed in a two-dimensional (2D) geometry using the Virtual Laser Plasma Lab (\texttt{VLPL}) code~\cite{CERNYellowReports_1_181,pukhov1999three}. The simulation domain is $100 \lambda_0$ and $50 \lambda_0$ in $x$ and $y$ direction ($\lambda_0 = 910 \text{nm}$ is the laser wavelength) with a spatial grid step of $0.02 \lambda_0 \times 0.05 \lambda_0$, respectively. The electromagnetic fields are updated with the X-dispersionless Maxwell solver \cite{Pukhov2019}, also known as RIP-solver. The Maxwell-solver requires $h_x = c \tau = 0.02 c / \omega_p$ with $h_x$ the longitudinal grid step, $\tau$ the time step and $\omega_p$ the non-relativistic plasma-frequency. A simulation runs for $120 T_0$ with $T_0 \approx 3.04 \textrm{fs}$ being the laser period.
The basic configuration is shown in Fig. \ref{fig:1_config}. 
The solid-state-target is located in the lower half of the simulation domain. The electron density is $505.55 n_{cr}$. Here, $n_{cr} \sim 1.35 \times 10^{21}$ is the critical density for the considered wavelength $\lambda_0$. Absorbing boundary conditions were chosen for the domain. The electrons are represented by four particles per cell.

Both lasers are linear polarized Gaussian beams with $a = a_0 \exp{\left(- (x-ct)^2/\tau^2-y^2/\sigma_y^2\right)}$
with $\tau = 8.240 T_0$ and $\sigma_y = 5.978 \lambda_0$. Each laser has a diameter of $5.978 \lambda_0$ and a length of $8.240 \lambda_0$.  The two lasers are grazing the target at different grazing angles $2.5^\circ \geq \Theta \geq 15^\circ$ ($1/72 \pi \geq \Theta \geq 1/12 \pi$) and are initialized $75 \lambda_0$ away from their point of incidence. Point of incidence of both laser propagation axes is at the centre of the upper edge of the target.
QED effects in \texttt{VLPL} are computed using the Monte Carlo method~\cite{elkina2011qed,Baumann2016}. Evaluated in the simulation data are the non-linear Compton scattering and the Breit-Wheeler process. 

Photon emission and the pair production have been implemented in QED sub-routines. Both routines are using the Monte Carlo approach and utilize the alternate model by Elkina et al. \cite{elkina2011qed}.
The first sub-routine which is called in a PIC loop is computing the Breit-Wheeler pair production that is run before updating the momentum of the particles in case the photon decays and does not require further calculations. First a random number $r_1 \epsilon [0,1]$ is decided which represents the possible energy of the electron of the pair. Afterwards, the probability rate is calculated with 
\begin{align}
\frac{dW_\textrm{pair}}{d\varepsilon_e} = \frac{\alpha m^2 c^4}{\hbar \varepsilon^2_\gamma} \left[ \int_x^\infty \textrm{Ai} \left( \xi \right) d \xi + \right. \\ \nonumber
+ \left. \left( \frac{2}{x} - \chi_\gamma \sqrt{x} \right) \textrm{Ai}^\prime \left( \chi \right) \right] .
\end{align} 	
A second random number decides whether the process occurs. The condition that needs to be fulfilled is
\begin{align}
r_2 < \left[ d W_\textrm{pair} / d \varepsilon_e \right] \varepsilon_\gamma \tau.
\end{align}
If the algorithm succeeds the photon gets deleted and electron and positron macro-particles are implemented in the simulation domain at the same place. The momentum of the pair abides the conservation of momentum. In the same manner the photon emission is simulated after the momentum update with its corresponding probability rates. Here, a photon is placed at the same location as the emitting particle and both particle momenta are calculated by the conservation of momentum.
\section{Results}
\label{sec:results}
\begin{figure*}
\centering
\includegraphics[scale=0.85]{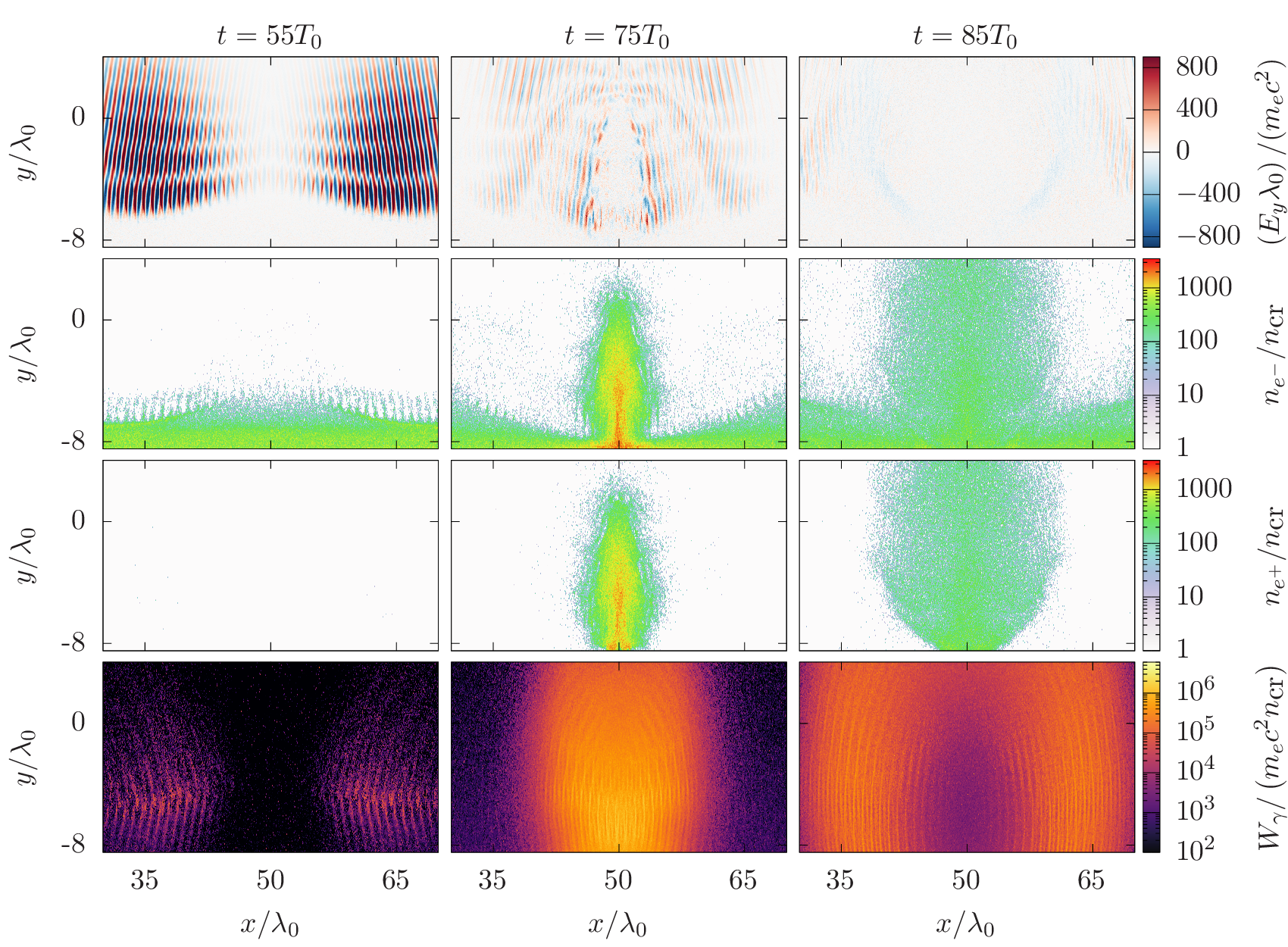}
\caption{$E_y$-component in dimensionless units (first row), electron density distribution in critical densities (second row), positron density in critical densities (third row), and energy density of emitted photons (fourth row) at different time instances in QED-PIC simulation with parameters $a_0 = 1200$ and $\theta = 15^{\circ}$.}
\label{fig:2_FD}
\end{figure*}

The first simulation presented uses lasers with an incident angle of $\Theta = 15^\circ$ and an $a_0 = 1200$. After the initialization, the lasers propagate along the surface extracting, capturing and accelerating electrons in the electromagnetic fields of the lasers (see Fig. \ref{fig:2_FD} first row). These electrons co-move with the laser along the surface (see figure \ref{fig:2_FD} second row $t = 55 T_0$). In the process the particles of the target emit photons, which can be seen in Fig. \ref{fig:2_FD} (fourth row). The energy density of the emitted photons is similarly structured to the propagating electromagnetic waves since the probability rate of the process is tied to the $\chi$-parameter, which includes the electromagnetic fields. \par

The trapped electrons and emitted photons produced by one laser beam collide with the counter-propagating laser beam and particles. In the interaction region, where both laser beams overlap, the non-linear Breit-Wheeler pair production becomes likely. At this point the $\chi$-parameter rises to a value of $9.65$ due to the strong field that a particle comes in contact within its rest frame. Photons decay in an electron-positron pair, which is represented by Fig. \ref{fig:2_FD} (third row) during the overlap (middle column) at $t=75 T_0$. An electron-positron plasma builds up in the region and expands outwards in the positive $y$-direction, where the target is not obstructing fields and particle dynamics, while reaching a higher peak density than the initial solid-state target. Pair production processes started to be recorded once the counter-propagating beam reaches the centre, since the fields near the surface are not sufficient to trigger the effect with the co-moving photons. 
In addition to the emitted photons by grazing the target, the collision of the extracted electrons and the respective laser triggers photon emission again, which fuels the electron-positron plasma. Once the field is partially absorbed by the electron-positron plasma, the new plasma is shielded by the remaining electromagnetic field (see Fig. \ref{fig:2_FD} first row) at $t = 85 T_0$. Several cycles of the emission of hard photons and the conversion of electron-positron pairs are observed leading to the electron-positron plasma by this QED cascade.

\begin{figure*}
\centering
\includegraphics[scale=0.75]{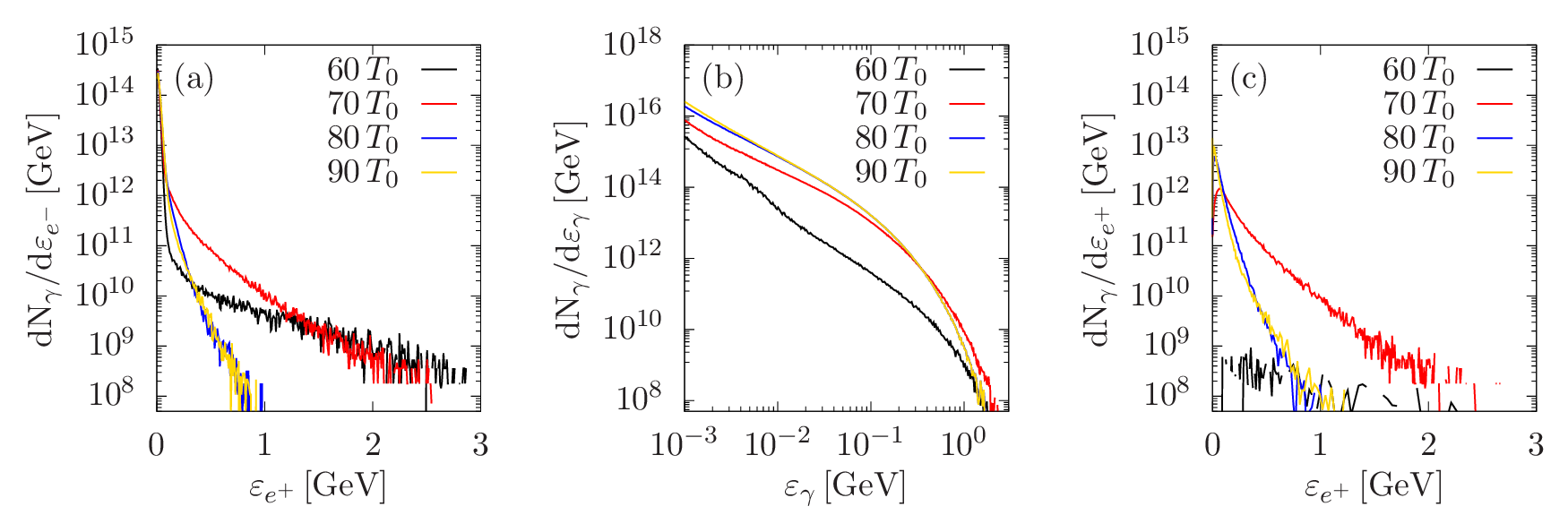}
\caption{Particle spectrum of electrons (a), photons (b) and positrons (c) at $t = 60 T_0, 70 T_0, 80 T_0$ and $90 T_0$. The highest peak of the laser beam reaches the centre of the surface at $t = 75 T_0$. Laser parameters are $a_0 = 1200$ and angle of incidence is $\Theta = 15^\circ$.}
\label{fig:3_time_eval}
\end{figure*}

Figure \ref{fig:3_time_eval} shows the spectra of electrons, positrons and $\gamma$-photons for four different time instances. The first displayed time at  $t = 60 T_0$ shows the energy spectrum after extracting and accelerating some electrons by the incident laser beams. Electrons, represented in subplot (a), are accelerated up to $3 \mathrm{GeV}$. Pair production at the early stage without interacting with the counter-propagating electromagnetic field (see subplot c) occurs only in a low number. Here, it can be seen, that the positron spectrum at $t =  60 T_0$ only shows some noise in the low energy region by a low number of pair production processes. This observation is in agreement with the rise of the electron-positron plasma shown earlier.

Continueing with the photon emission, $\gamma$-photons are emitted at two points in the configuration. First, photon emission takes place once the laser beam comes into contact with the target and then continuously emits photons while grazing the surface, which can be seen in figure \ref{fig:3_time_eval}b where the energy density of the photons shows the characteristic photon spectrum at $t = 65 T_0$. Second, the accelerated and extracted electrons collide with the counter-propagating beam and radiate high-energy due to the stronger electromagnetic field perceived in the electrons rest frame. In the time instance $t = 70 T_0$ and forward the number of photons increases by three orders of magnitudes . During the overlap the photon spectrum enhances the high-energy photon emission, since photons with an energy between $10 \textrm{MeV}$ and $1 \textrm{GeV}$ are emitted.

As the electron beam interacts with the counter-propagating laser beam the number of positrons increases. This indicates that the probability for pair production processes became more likely. Both the electron and positron spectrum roughly coincide.   
With the counter-propagating laser beam reaching the point of incidence the maximum recorded electron energy drops to $\sim 1 \mathrm{GeV}$. The electrons lose their energy due to the interaction with the counter-propagating laser. We observe this as the increase of low-energy electrons in the spectrum. At this point more $\gamma$-photons are emitted that further produce electron-positron pairs. This is indicated by the positron spectrum in subplot (c) at $t = 80 T_0$. Again at $t = 90 T_0$ the positron spectrum shifts towards low-energy positrons similar to electrons.

\begin{figure*}
\centering
\includegraphics[scale=0.75]{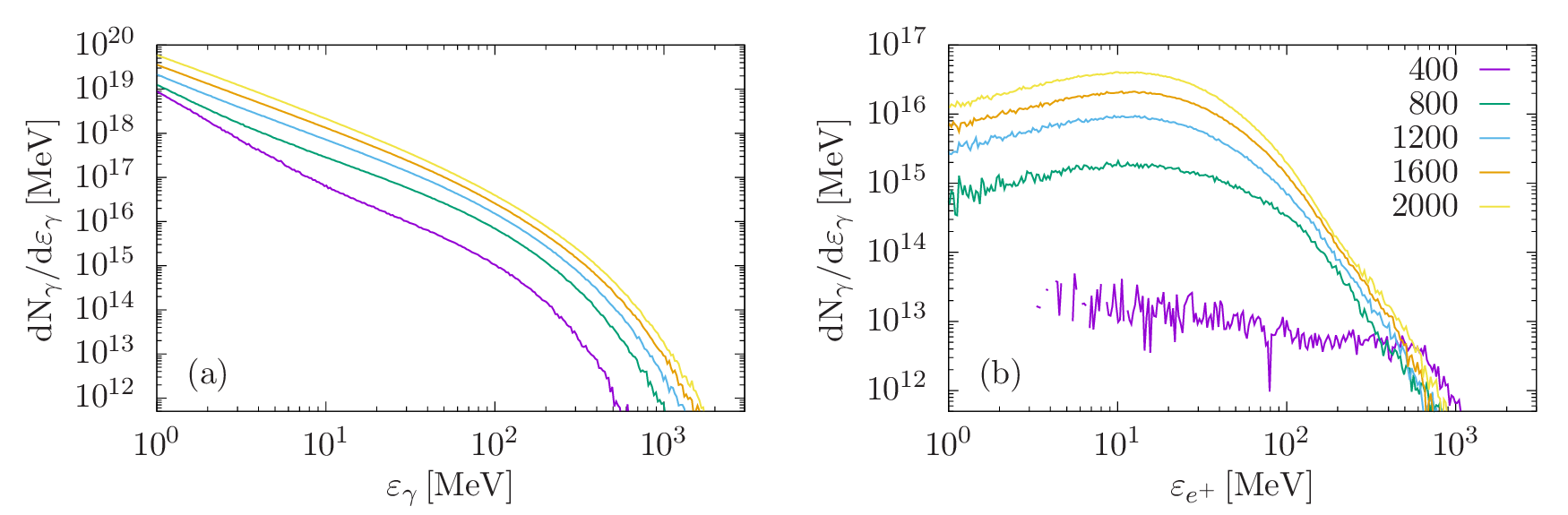}
\caption{Double logarithmic spectrum of emitted photons (a), and logarithmic spectrum of positrons (b) after the overlap of the high-intensity lasers for different $a_0$ pulses. Angle of incidence for both configuration is $\Theta = 15^\circ$.}
\label{fig:4_spectrum_diff_a0}
\end{figure*}

In a next step, the influence of the laser parameters will be discussed.
Figure \ref{fig:4_spectrum_diff_a0} shows the photon spectrum and positron spectrum for different laser amplitudes ranging from 400 to 2000. In general, increasing the energy of the laser beam boosts the secondary particle spectra. Additionally, the cutoff energy of the photon drifts to a higher value by increasing $a_0$. In the special case of $a_0 = 400$ the characteristic spectrum of positrons is not reproduced since the statistic is insufficient and the electromagnetic fields are not strong enough to develop the positron spectrum. Only by reaching an $a_0 \sim 800$ pair processes are sufficiently witnessed and an electron-positron plasma builds up.

\begin{figure}
\centering
\includegraphics[scale=1.0]{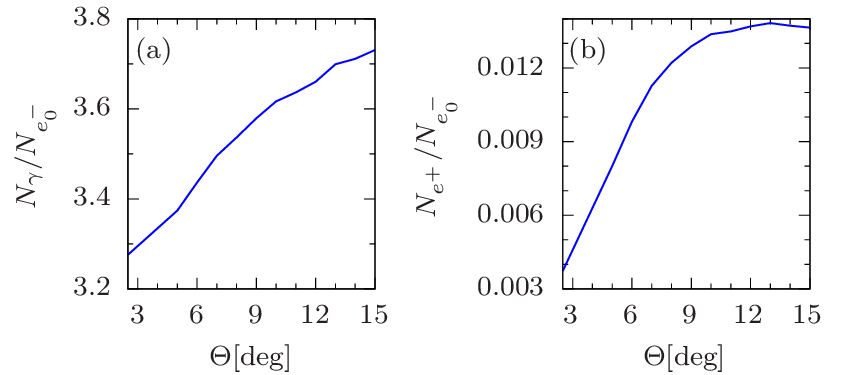}
\caption{Ratio of emitted photons to the initial number of electrons (a), ratio of electron-positron pairs to the initial number of electrons (b)}.
\label{fig:5_fraction}
\end{figure}

The other laser parameter in the proposed configuration is the angle of incidence $\Theta$. In a second simulation series the angle has been varied between $2.5 - 15 \left[\textrm{deg} \right]$ while maintaining the dimensionless vector amplitude at $a_0 = 800$. Figure \ref{fig:5_fraction} shows the ratios for emitted photons (subplot a) and positrons (subplot b) per initial electron. In general, fractions of secondary particles increase with a larger angle $\Theta$. Further, positrons of the pair production per initial electrons are maximized at an angle of $\sim 10^\circ$, whereas photons remain to increase with bigger angles. A possibility for this observation may be the energy loss of the electromagnetic fields. The fields are absorbed by the electron-positron plasma in the interaction region. When the laser energy is depleted, the pair production ceases and the ratio of positrons per initial electrons is maximized. While this is the case for pair production, photons may still be emitted with a weaker field. Fraction of photons per initial electron continue to rise after an incident angle of $10^\circ$. \par

\begin{figure}
\centering
\includegraphics[scale=1.0]{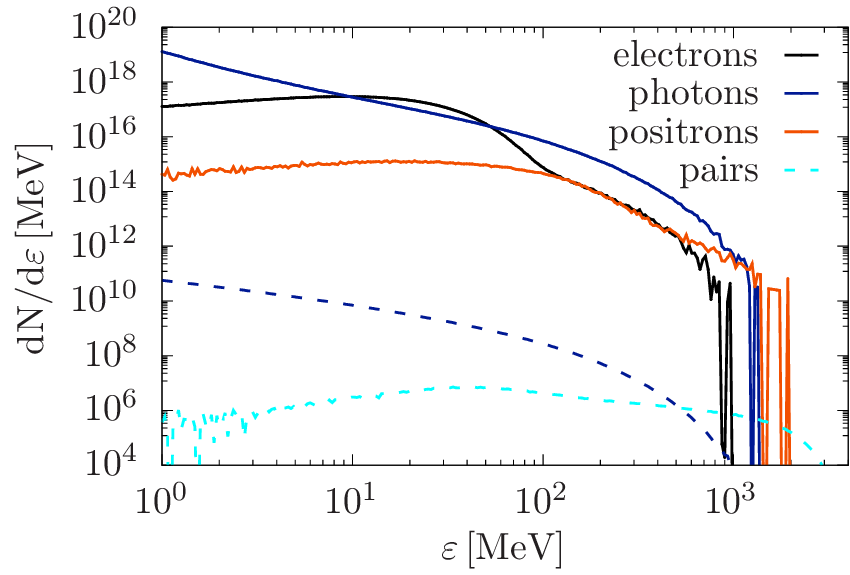}
\caption{Double logarithmic spectrum of secondary particles. Solid line represents the configuration of grazing a solid-state target; dashed line is the seeded vacuum configuration. Spectrum after the overlap of the high-intensity lasers at $t = 90 T_0$. Angle of incidence for both configuration is $\Theta = 12^\circ$ and $a_0 = 800$. Additionally, the cyan line represents the electron and positron spectrum, since both spectrum are equal due to the small number of seed electrons.}
\label{fig:6_compare_vacuum}
\end{figure}

The previous results showed that QED effects were observed in the proposed configuration. In a final step, the configuration will be compared to a seeded vacuum cascade \citep{tamburini2017laser,PhysRevLett.105.080402,doi:10.1063/1.5022640}. Setting up a configuration for a vacuum cascade appears to be simpler, therefore it is reasonable to compare both configurations. A seeded vacuum may resemble an imperfect vacuum, where a small impurity remains after trying to create a vacuum. Seed electrons are necessary to initiate QED effects in the code. Fig. \ref{fig:6_compare_vacuum} shows a comparison between the proposed configuration and the seeded vacuum within the secondary particle spectra for lasers with $a_0 = 800$ and $\theta = 12^\circ$. $72$ seed electrons are initialized in two cells where the propagation axis of both lasers intersect. The density in those cells is $n_{cr} = 7.39 \times 10^{-7}$. Here, the solid-state-target emits many order of magnitudes of photons more than the seeded vacuum cascade. While emitting less photons the vacuum cascade manages to accelerate the positrons created by pair production process to an energy of $\sim 3 \textrm{GeV}$ seen in the increased cutoff energy. The proposed configuration still outperforms the vacuum scenario in the yield of pairs. However, the maximum photon energy achieved is $\sim 1400 \textrm{MeV}$.

\section{Conclusion}
\label{sec:conclusion}
In this paper, the interaction of two high-intensity lasers and a solid-state target was studied in the framework of PIC simulations. Focus was placed on QED processes where both lasers overlap and interact with extracted and accelerated electrons. The large number of the extracted electrons escalates into a QED cascade creating an electron-positron plasma once the laser intensity was sufficiently high. Comparing the plasma with a seeded vacuum cascade demonstrated that using a target outperforms an imperfect vacuum.

Further, increasing the angle of incidence reached an upper limit on produced pairs by QED effects on the electron-positron plasma and higher lasers intensities showed, that a certain laser intensity is necessary to trigger pair production processes.

Additional studies should be performed on different potential target materials and other forms to enhance the achieved quantum parameter or increase the number of QED processes.

In the near future where higher intensities are reached this configuration may be replicated experimentally and help to achieve even not yet experimentally explored regimes like the fully nonpertubative regime.
\section*{Acknowledgments}
One of the authors (M.F.) is thankful for valuable discussion with Christoph Baumann. This work has been funded by the Deutsche Forschungsgemeinschaft (DFG) under project number 430078384. The authors gratefully acknowledge the Gauss Centre for Supercomputing e.V. (\url{www.gauss-centre.eu}) for funding this project (qed20) by providing computing time through the John von Neumann Institute for Computing (NIC) on the GCS Supercomputer JUWELS\cite{JUWELS} at J\"ulich Supercomputing Centre (JSC).

\bibliographystyle{unsrt}
\bibliography{arxiv_grazing_filipovic}  

\end{document}